# A Network Control Theory Approach to Longitudinal Symptom Dynamics in Major Depressive Disorder

Tim Hahn [1*†], Hamidreza Jamalabadi [2*], Daniel Emden [1], Janik Goltermann [1], Jan Ernsting[1,3], Nils R. Winter [1], Lukas Fisch [1], Ramona Leenings [1,3], Kelvin Sarink [1], Vincent Holstein [1], Marius Gruber [1], Dominik Grotegerd [1], Susanne Meinert [1,4], Katharina Dohm [1], Elisabeth J. Leehr [1], Maike Richter [1], Lisa Sindermann [1], Verena Enneking [1], Hannah Lemke [1], Stephanie Witt [5], Marcella Rietschel [5], Katharina Brosch [6], Julia-Katharina Pfarr [6], Tina Meller [6], Kai Gustav Ringwald [6], Simon Schmitt [6], Frederike Stein [6], Igor Nenadic [6], Tilo Kircher [6], Bertram Müller-Myhsok[7], Till F.M. Andlauer [7], Jonathan Repple [1], Udo Dannlowski [1], Nils Opel [1,8]

[1] Institute for Translational Psychiatry, University of Münster

[2] Department of Psychiatry and Psychotherapy, University of Tübingen

[3] Faculty of Mathematics and Computer Science, University of Münster

[4] Institute for Translational Neuroscience, University of Münster

[5] Department of Genetic Epidemiology, Central Institute of Mental Health, Faculty of Medicine Mannheim, University of Heidelberg, Mannheim, Germany

[6] Department of Psychiatry and Psychotherapy, Phillips University Marburg

[7] Max-Planck-Institute of Psychiatry, Munich

[8] Interdisciplinary Centre for Clinical Research, University of Münster

* these authors contributed equally to this work

†Corresponding author:

Tim Hahn, Ph.D., Department of Psychiatry, University of Münster, Germany

Albert-Schweitzer-Campus 1, D-48149 Münster

Phone:  +49 (0)2 51 / 83 – 56625, Fax:    +49 (0)2 51 / 83 – 56612, E-Mail: HahnT@wwu.de

**Abstract**

**Background**: The evolution of symptoms over time is at the heart of understanding and treating mental disorders. However, a principled, quantitative framework explaining symptom dynamics remains elusive. Here, we propose a *Network Control Theory of Psychopathology* allowing us to formally derive a theoretical control energy which we hypothesize quantifies resistance to future symptom improvement in Major Depressive Disorder (MDD). We test this hypothesis and investigate the relation to genetic and environmental risk as well as resilience.

**Methods**: We modelled longitudinal symptom-network dynamics derived from N=2,059 Beck Depression Inventory measurements acquired over a median of 134 days in a sample of N=109 patients suffering from MDD. We quantified the theoretical energy required for each patient and time-point to reach a symptom-free state given individual symptom-network topology ($E_0$) and 1) tested if $E_0$ predicts future symptom improvement and 2) whether this relationship is moderated by Polygenic Risk Scores (PRS) of mental disorders, childhood maltreatment experience, and self-reported resilience.

**Outcomes**: We show that $E_0$ indeed predicts symptom reduction at the next measurement and reveal that this coupling between $E_0$ and future symptom change increases with higher genetic risk and childhood maltreatment while it decreases with resilience.

**Interpretation**: Our study provides a mechanistic framework capable of predicting future symptom improvement based on individual symptom-network topology and clarifies the role of genetic and environmental risk as well as resilience. Our control-theoretic framework makes testable, quantitative predictions for individual therapeutic response and provides a starting-point for the theory-driven design of personalized interventions.

**Funding**: German Research Foundation and Interdisciplinary Centre for Clinical Research, Münster

**1. Introduction**

In psychiatry, symptoms not only aid diagnosis, but define disorders.[1] Consequently, the evolution of symptoms over time – i.e., symptom dynamics – is at the heart of understanding and treating mental disorders. For example, clinicians make extensive use of symptom assessment over time to monitor patients' disease trajectories. Similarly, therapeutic interventions can be understood as an effort to control symptom dynamics. Despite this central role in clinical practice, a quantitative framework providing mechanistic insight into symptom dynamics and control remains elusive.

To address this issue, we first consider the *Network Theory of Psychopathology* (NTP).[2] It posits that mental disorders can be understood as systems of interdependent symptoms – so-called symptom-networks. Within this theory, mental disorders are not merely characterized by symptoms, but are the consequence of a dynamical system of symptoms evolving over time.[3] The central tenet of NTP is the notion that symptoms can affect symptoms. For example, a symptom-network can stabilize or reinforce itself to hamper recovery or increase severity. Likewise, treating one symptom may decrease other symptoms. After more than a decade of research, overwhelming evidence supports the interdependence of symptoms – i.e., a non-random symptom-network topology.[2–5]

While NTP has not been formalized and the predictive power of the approach remains in question,[2,6] uncovering non-random symptom-network topology, in our view, provides the key for a more principled theoretical approach: Here, we posit that symptom dynamics can be modeled using *Network Control Theory* which offers a mathematical framework for analyzing and, ultimately, controlling the evolution of dynamical systems.[7–9] Specifically, we suggest treating psychopathology analogous to a dynamical system in the physical world. We propose to model symptom dynamics via a linear time-invariant control system relating 1) a patient's symptoms over time, 2) the symptom-network topology derived from NTP as outlined above, and 3) an actual or theoretical intervention on specific symptoms (for a comprehensive introduction to *Control Theory*, see [10]; see *Methods* for details). This view not only allows us to model symptom dynamics based on symptom-network topology, but to frame therapeutic interventions as the manipulation of a dynamical system of symptoms so that it evolves to follow a particular trajectory through its state space – e.g. towards a symptom-free state. Taken together, our *Network Control Theory* extension of NTP yields a quantitative theory of symptom dynamics and intervention from which we can derive concrete hypotheses.

This allows us to address the fundamental question of symptom trajectory prediction in Major Depressive Disorder (MDD; for previous attempts to predict future symptom states based on NTP, see *Supplementary Introduction*). To this end, we capitalize on recent advances in *Network Control Theory* which enable the quantification of the minimum theoretical energy required to move the system's state from one location to another – corresponding to different symptom configurations and severity. Treating psychiatric disorders as dynamical systems, this energy represents the resistance to symptom change. Importantly, the energy required to change symptoms depends not only on the symptoms at a given time-point, but on individual symptom-network topology.[11–13]

To test the explanatory power of our framework, we model longitudinal symptom-network dynamics derived from N=2,059 Beck Depression Inventory (BDI) measurements acquired over a median of 134 days in a sample of N=109 patients suffering from MDD using *Network Control Theory*. Then, we quantify the theoretical energy required for each patient at each time-point to reach a symptom-free state given his/her current symptoms and individual symptom-network topology ($E_0$, cf. Equation 2 in *Methods*). Within the framework of the linear time-invariant control system formulation, $E_0$ quantifies the system's current resistance to be driven to a symptom-free state. Thus, we hypothesize that $E_0$ at a given time-point is negatively associated with BDI improvement (i.e., relative movement towards a symptom-free state) at the next time point. For a schematic overview of the analysis conducted for each patient, see Figure 1.

As NTP and our *Network Control Theory* extension are based solely upon symptom assessments, our quantification of $E_0$ is based only upon those symptoms captured by BDI measurements and their network structure. Thus, we assume that the hypothesized coupling between $E_0$ and future BDI improvement is affected by additional factors not considered in our model. In particular, the current approach does not speak directly to the role of biological or environmental factors such as genetic and environmental risk known to be relevant in MDD. To bridge this gap, we investigated the effects of genetic risk of mental disorders – namely for MDD and Bipolar Disorder – and childhood maltreatment, representing prominent genetic and environmental risk factors previously associated with MDD.[14,15] Complementing this focus on individual risk, we also investigated resilience – i.e., the predispositions helping to maintain a stable level of psychological functioning in the face of traumatic stress.[16] As a major individual resource for patients, resilience has previously been suggested as a protective factor in MDD.[17]

## Research in context

### Evidence before this study

The *Network Theory of Psychopathology* (NTP) posits that mental disorders can be conceptualized as causal systems of mutually reinforcing symptoms. While this has greatly enriched our understanding of mental disorders, a mathematically principled, quantitative framework modeling symptom-network dynamics remains elusive. To assess the existing conceptual and empirical work relating to symptom-network dynamics in Major Depressive Disorder, we searched PubMed and Web of Knowledge databases in English for work published between Jan 1, 2000, and Nov 15, 2020, using the search terms ((MDD) OR (major depression) OR (major depressive disorder) OR (depression) OR (unipolar depression)) AND ((network theory of psychopathology) OR (symptom network) OR (symptom dynamics)). In short, overwhelming evidence supports NTP's central tenet of the interdependence – i.e., a non-random network topology – of symptoms. However, the theory does not provide a principled means to guide interventions and its predictive power with regard to symptom dynamics remains in question.

### Added value of this study

To our knowledge, this is the first work to suggest an overarching, comprehensive theory of symptom dynamics and its control. By connecting the *Network Approach to Psychopathology* with *Network Control Theory*, we can leverage a rich mathematical framework to analyze the influence of a control signal on the dynamics of interconnected systems. This allows us to formally derive a control energy construct ($E_0$) which theoretically quantifies resistance to reach a symptom-free state. Empirically, we show that $E_0$ is indeed empirically associated with future symptom improvement in a longitudinal study. Extending the current symptom-focused view, we additionally show that genetic risk of mental disorders, childhood maltreatment experience, and resilience moderate this newly found coupling of $E_0$ and symptom improvement.

**Implications of all the available evidence**

Our approach connects a major theory of mental disorders to the mathematical study of system control. By showing that a theoretical energy construct based formally upon symptom-network topology is indeed associated with future symptom improvement over several months, we hope to lay the groundwork for a *Network Control Theory of Psychopathology*. Further, we bridge the gap between our approach and genetic and environmental factors thus far not considered in NTP research. Finally, our approach makes testable, quantitative predictions for individual therapeutic response to be investigated in future studies (see *Discussion* for details) and provides a starting point for the theory-driven design of personalized interventions directly on specific symptoms.

**2. Methods**

<u>Participants</u>

At the time of data analysis, 450 cases of MDD had been recruited at baseline in the ReMAP study, from 113 of which eight or more BDI measurements were available. In four patients, the estimation of $E_0$ did not converge due to insufficient variance in BDI scores. Thus, data from N=109 patients (n=85 female) between 18 to 64 (mean=40.6) years of age entered the analysis. Patient recruitment and data assessment were conducted by the Institute of Translational Psychiatry, Department of Psychiatry at the University of Münster. Patients with severe, moderate, mild or (partially) remitted MDD episodes were included irrespective of current treatment (see Supplementary Tables S1 and S2). A structured clinical interview for DSM-IV (SCID-I) was conducted with each participant in order to assess current and lifetime psychopathological diagnoses.[18] Patients either fulfilled the DSM-IV criteria for an acute major depressive episode or had a lifetime history of a major depressive episode. The study design, including recruitment procedures and technical feasibility have been described in previous work (see Supplementary Material).[19,20]

Psychometric Assessment of Depressive Symptoms, Childhood Maltreatment and Resilience

Following previous work in NTP[21], we assessed depressive symptoms using the German version[22] of the Beck Depression Inventory (BDI).[23] Childhood maltreatment was measured using the Childhood Trauma Questionnaire (CTQ)[24] in its German version.[25] CTQ data was available from N=93 patients. We assessed resilience via the German version[26] of the Resilience Scale.[27] Resilience Scale data was available from N=68 patients.

Genotyping and calculation of Polygenic Scores

Genotyping was conducted using the PsychArray BeadChip, followed by quality control, population substructure analyses, and imputation, as described previously.[28,29] Imputed genetic data were available for n=60 individuals. For the calculation of polygenic risk scores (PRS; [30]) from genome-wide association studies on major depression[31] and bipolar disorder[32], SNP weights were estimated using the PRS-CS method [33] with default parameters (for details, see *Supplementary Methods*). In these analyses, three ancestry components were used as covariates.

Longitudinal Assessment of Symptom Dynamics

To obtain longitudinal symptom dynamics, we acquired BDI measurements via the smartphone-based *Remote Monitoring Application in Psychiatry* (ReMAP).[19,20,34] Importantly, extensive validation of BDI assessment via ReMAP indicates overall high comparability between smartphone-based and stationary BDI scores (ICC=.921, p<.001).[20]

Symptom-Network Topology

To obtain the symptom-network topology (i.e. the correlational structure of symptoms over time) as known from NTP, we computed the partial correlation between each pair of items in the BDI while controlling for all other symptoms for each individual separately. To ensure a reliable estimate, a minimum of eight BDI measurements had to be available. This procedure resulted in a 21x21 symmetrical matrix quantifying undirected association between symptoms for each patient.[5] To account for the ordinal scale of BDI items (ranking 0, 1, 2 or 3 for increasing symptom severity), we used Kendall's correlation coefficient. Note that we prefer Kendall over Spearman rank correlation due to its improved robustness at low sample sizes.[35] We retained significant correlations only and controlled for multiple comparisons by calculating the false discovery rate with a false-positive rate of 0.05 over all unique, non-trivial correlations.[36]

Network Control Theory

The aim of network control is to drive a dynamical system towards a desired state by influencing a selection of input nodes.[7] Following previous work [37], we assume the system to follow a noise-free linear time-invariant model given by

$$\dot{x} = Ax(t) + Bu(t) \tag{1}$$

where x(t) defines the state of the system at time t, A represents the interaction matrix, B is the input matrix, and u(t) corresponds to the input signal (for a comprehensive introduction to *Control Theory*, see [10]). In case of symptom dynamics, x(t) corresponds to a patient's current symptoms, A is the symptom-network topology derived from NTP as outlined above, B specifies which symptoms can theoretically be targeted by an intervention, and u(t) corresponds to an actual or theoretical intervention on specific symptoms. A rigorous mathematical formulation of network controllability in brain networks can be found in [37].

Based on this, we computed the energy required to move from the current symptom-state ($BDI_t$) to a symptom-free state in which all BDI items have a value of 0 as follows:

$$E_0 = u(t)^T u(t) \tag{2}$$

where u(t) is the solution to the following optimal control problem:

$$\min_u \int_0^T \left(x_T - x(t)\right)^T S\left(X_T - x(t)\right) + \rho u(t)^T u(t),$$

$$\text{(3)} s.t. \dot{x} = Ax(t) + Bu(t), x(0) = BDI_t \text{ and } x(T) = 0$$

$$\text{(4)}$$

where $T$ and $\rho$ are free parameters quantifying the time to reach to symptom-free state and the relative importance of cost terms in equation (3). Following Gu et al.[12], we define the step size to 0.001 and T = 1. As we are interested in moving the current state $BDI_t$ with minimum external energy (i.e. actual or theoretical intervention) to the symptom-free state, we set S = 0, $\rho$=1, and B to the identity matrix to keep all 21 symptom nodes as potential intervention bases. To solve the optimal control equations (3) and (4), we follow Gu et al. and use a customized version of the code available from [12].

Statistical Analysis

To test our main hypothesis, we computed, for each patient, the partial correlation based on Spearman's correlation coefficient between the energy required to move to a symptom-free state at time t (i.e., $E_{0t}$) and symptom change at the next time point (i.e. $BDI_{t+1} - BDI_t$) while 1) controlling for overall symptom severity itself (i.e. $BDI_t$) to account for autocorrelation of symptom severity and avoid ceiling or floor effects and 2) the number of days between measurements (i.e., $days_{t+1} - days_t$) to account for differing gaps between measurements. Then, we took the mean of the Fisher transformed correlation coefficients across all patients and computed the p-value for the correlation based on the one-tailed Student's t-distribution with n-2 degrees of freedom. The predictive performance of a linear regression model predicting BDI improvement from $E_0$ and previous BDI, respectively, was assessed with Mean Absolute Error (MAE) via leave-one-out cross-validation using scikit-learn.[38]

To test the effects of PRS, CTQ, and resilience, we employed an ANCOVA approach with age, gender, and number of BDI measurements as covariates (all results are provided with two-tailed p-values). In the analyses of PRS, three ancestry components were used as covariates.

Role of the funding source

The funders had no role in designing the study; collection, management, analysis, or interpretation of data; writing of the report; nor the decision to submit the report for publication. The authors had the final responsibility for the decision to submit for publication.

3. Results

Confirming previous findings of non-random network topology, we show significant interdependence between symptoms across participants and time-points (p<0.05, FDR-corrected; Supplementary Figure S3). Testing our main hypothesis, we assessed whether $E_0$ at a given time-point is associated with symptom change at the next time point (Figure 1). As hypothesized, $E_0$ was negatively associated with symptom change at the next time point (mean Spearman r(107)=-0.26; 95% CI [-.40 − -.10]; p=.003; one-tailed) across all participants, indicating that a higher theoretical energy required to reach a symptom-free state leads to less improvement (or worsening) at the next time point. Likewise, we could predict the BDI improvement from $E_0$ with an MAE of 2.72 in a leave-one-out cross-validation framework, outperforming a prediction based on BDI alone (MAE=3.34; p<.001).

While evidence thus confirms our hypothesis, we observe substantial individual variance indicating differences in how well $E_0$ is associated with future symptom improvement (Supplementary Figure S4). Bridging the gap between biological and environmental factors, we first tested whether this variance in coupling between $E_0$ and symptom improvement can be explained by genetic risk of mental disorders. We show that PRS quantifying the genetic risk of MDD are negatively associated with the coupling between $E_0$ and symptom improvement ($F(1,52)=7.58$, $p=0.008$). We observed a similar effect for PRS quantifying the genetic risk of bipolar disorder ($F(1,52)=4.76$, $p=0.034$), implying that the absolute coupling between $E_0$ and symptom improvement is stronger in individuals with higher risk of affective disorders (Figure 2, top row). Second, we tested whether the effect observed for genetic risk could also be found for environmental risk. We show that the extent of childhood maltreatment as measured by the Childhood Trauma Questionnaire (CTQ) also has an above-chance effect in the same direction as genetic risk ($F(1,88)=5.90$, $p=0.017$; Figure 2, bottom row).

To complement this focus on individual risk, we finally investigated resilience as measured by the Resilience Scale. In contrast to genetic and environmental risk, resilience is positively associated with the coupling between $E_0$ and symptom improvement ($F(1,63)=9.71$, $p=0.003$; Figure 2, bottom row). In addition, both subscales show positive associations (*Acceptance of Self and Life*: $F(1,63)=10.45$, $p=0.002$; *Personal Competence*: $F(1,63)=3.85$, $p=0.054$), i.e. lower coupling of $E_0$ and future symptom improvement for more resilient patients.

## 4. Discussion

In this work, we drew upon *Network Control Theory* to model longitudinal symptom dynamics in MDD. This approach offers a mathematical framework for analyzing and controlling the evolution of dynamical systems.[7–9] We derived a theoretical control energy, $E_0$, which quantifies the energy required to reach a symptom-free state. Testing the predictive power of our approach, we show that $E_0$ is indeed negatively associated with future symptom improvement. The finding that higher theoretical energy required to reach a symptom-free state led to less improvement (or worsening) at the next time point, supports the notion that symptom dynamics can – at least in part – be understood analogous to physical dynamical systems within the formal framework of *Network Control Theory*.

Our findings open the door towards a quantitative understanding of symptom dynamics. For example, previous work from NTP suggested that more strongly connected symptom networks ought to be less likely to change over time.[39] While sharing the fundamental idea that symptom-network topology affects future symptom trajectories, *Network Control Theory* clarifies how different symptom-states differ quantitatively with regard to the energy required to reach them. Thus, individual differences in network topology – not merely stronger connections between symptoms – determine resistance to symptom change in our approach. Also, resistance to change is not the same in every direction of symptom-space, but certain trajectories are more easily accessible than others.

Of practical relevance, *Network Control Theory* provides a straightforward way to calculate which symptoms should be addressed in which order to efficiently reach a symptom-free state. Whether treating symptoms identified by this so-called control-node analysis[10] leads to quicker recovery can be directly tested in future intervention studies. Importantly, this view also entails that therapeutic interventions should be custom-tailored to the patient's individual symptom-network topology which might thus serve as a promising approach to patient stratification (for additional analyses identifying optimal symptoms for intervention, see *Supplementary Results*)

Despite the insights and opportunities that may emerge from a *Network Theory of Psychopathology*, it does not speak directly to the role of biological or environmental factors. Bridging this gap, we reveal that the coupling between $E_0$ and future symptom-change increases with higher genetic risk for MDD and bipolar disorder as well as with childhood maltreatment experience.

In contrast, resilience appears to decrease this relationship, rendering individuals less affected by their symptom-network topology. In other words, higher genetic or environmental risk may be interpreted as less flexibility to deviate from the symptom dynamics dictated by symptom-network topology while higher resilience allows patients to deviate from the trajectory outlined by the interrelation of symptoms. Opening new vistas onto risk and prevention, this is compatible with the notion that higher genetic or environmental risk may render patients unresponsive to interventions due to a stronger dependence on their symptom-network structure. Likewise, the outcome of more resilient patients may be less affected by their symptom-network structure, making them more responsive to intervention. This interpretation is consistent with previous evidence linking high genetic and environmental risk to unfavorable outcomes (see Supplementary Discussion for details).[40,41]

Although we provide a theoretical framework and initial, encouraging evidence, studying symptom dynamics in the context of *Network Control Theory* remains challenging. Crucially, the estimation of symptom-network topology requires longitudinal symptom assessment over extended periods of time. Here, we relied upon data from the ReMAP study which combines longitudinal, smartphone-based symptom assessments with the acquisition of genetic, psychometric, and neuroimaging data.[19] To foster our understanding of symptom dynamics, similar initiatives will have to augment existing large-scale studies with longitudinal measurements. Alternatively, making the myriad of symptom assessments conducted as part of the daily clinical routine accessible would boost this new field beyond what any single study or consortium could achieve. While connecting these assessments to genetic, psychometric, and neuroimaging data will realize the full potential of the approach, logistic and data protection issues have to be addressed. Importantly, a *Network Control Theory of Psychopathology* offers the unique opportunity to link clinical practice with genetics and neuroscience within a common, mathematically principled framework. For an in-depth discussion of technical limitation and methodological challenges for future research, we refer the reader to the *Supplementary Discussion*.

In summary, we put forward a mechanistic framework rooted in *Network Control Theory* and show that future symptom improvement is associated with the theoretical energy required to reach a symptom-free state. Importantly, we extend this approach to clarify the role of genetic and environmental risk as well as resilience. Within the broader domain of Computational Psychiatry, our *Network Control Theory* extension of NTP provides a principled framework which – with its direct implications for the study of symptom dynamics – makes testable, quantitative predictions regarding individual therapeutic response and outcome. Crucially, it provides a starting point for the theory-driven design of personalized interventions which we hope will spark future research towards a more comprehensive, quantitative *Network Control Theory of Psychopathology*.

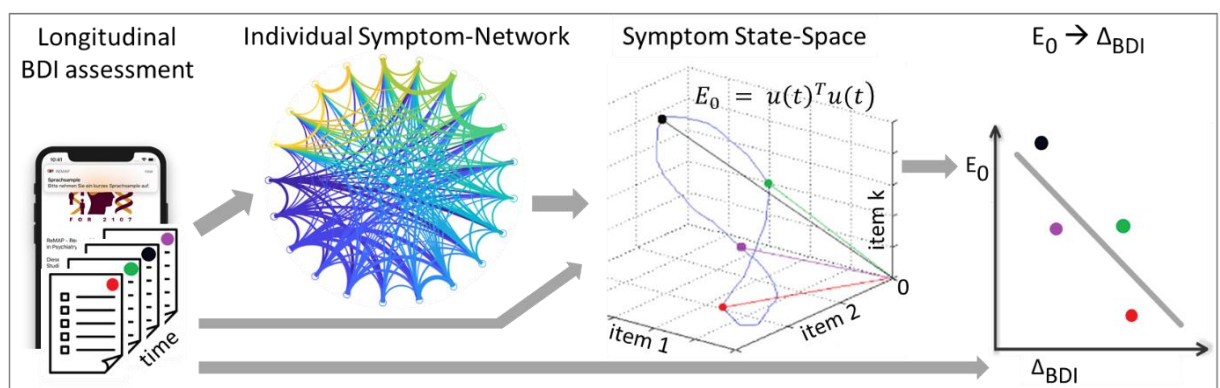

Figure 1. Single-patient analysis overview. From longitudinal BDI data (1st panel from left), we computed individual symptom-network topology (2nd panel). Combining symptom-network topology and current symptom-state, we quantified the theoretical energy required at each time-point to reach a symptom-free state ($E_0$, 3rd panel). Finally, we estimated the association between $E_0$ at each time-point and BDI improvement at the next time-point ($E_0 \rightarrow \Delta_{BDI}$).

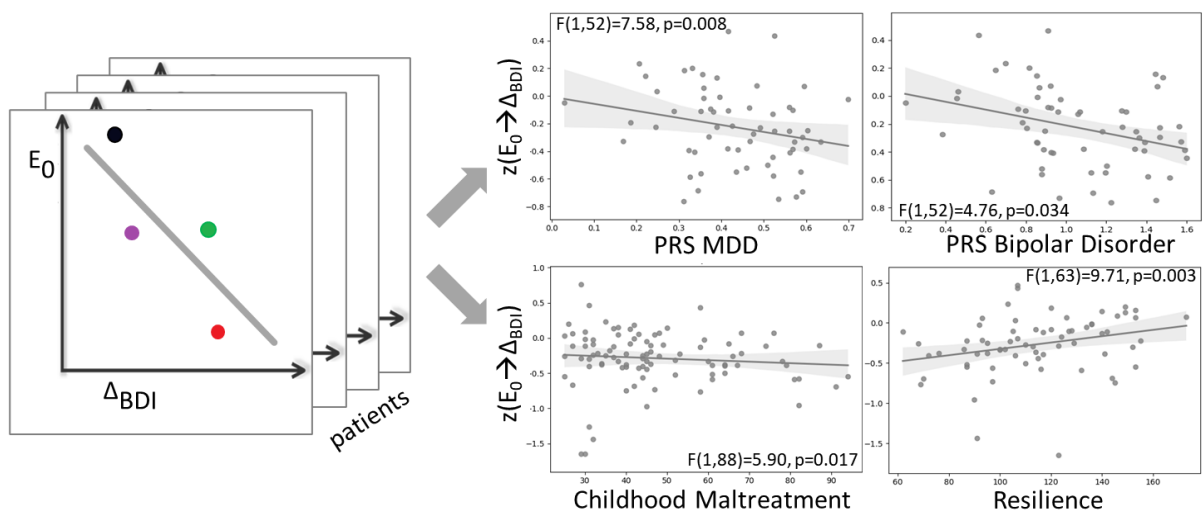

Figure 2. Association between the coupling of $E_0$ and BDI improvement at the next time-point ($E_0 \rightarrow \Delta_{BDI}$, cf. Fig. 1) for each patient and genetic risk of MDD and Bipolar Disorder (top row) as well as childhood maltreatment (bottom row, left panel) and resilience (bottom row, right panel). PRS: Polygenic Risk Score; MDD: Major Depressive Disorder.

Funding

This work was funded by the German Research Foundation (DFG grants HA7070/2-2, HA7070/3, HA7070/4 to TH) and the Interdisciplinary Center for Clinical Research (IZKF) of the medical faculty of Münster (grants Dan3/012/17 to UD , MzH 3/020/20 to TH and SEED 11/19 to NO) as well as the "Innovative Medizinische Forschung" (IMF) of the medical faculty of Münster (Grants OP121710 to NO and TH).

The MACS dataset used in this work is part of the German multicenter consortium "Neurobiology of Affective Disorders. A translational perspective on brain structure and function", funded by the German Research Foundation (Deutsche Forschungsgemeinschaft DFG; Forschungsgruppe/Research Unit FOR2107). Principal investigators (PIs) with respective areas of responsibility in the FOR2107 consortium are: Work Package WP1, FOR2107/MACS cohort and brainimaging: Tilo Kircher (speaker FOR2107; DFG grant numbers KI 588/14-1, KI 588/14-2), Udo Dannlowski (co-speaker FOR2107; DA 1151/5-1, DA 1151/5-2), Axel Krug (KR 3822/5-1, KR 3822/7-2), Igor Nenadic (NE 2254/1-2), Carsten Konrad (KO 4291/3-1). WP2, animal phenotyping: Markus Wöhr (WO 1732/4-1, WO 1732/4-2), Rainer Schwarting (SCHW 559/14-1, SCHW 559/14-2). WP3, miRNA: Gerhard Schratt (SCHR 1136/3-1, 1136/3-2). WP4, immunology, mitochondriae: Judith Alferink (AL 1145/5-2), Carsten Culmsee (CU 43/9-1, CU 43/9-2), Holger Garn (GA 545/5-1, GA 545/7-2). WP5, genetics: Marcella Rietschel (RI 908/11-1, RI 908/11-2), Markus Nöthen (NO 246/10-1, NO 246/10-2), Stephanie Witt (WI 3439/3-1, WI 3439/3-2). WP6, multi method data analytics: Andreas Jansen (JA 1890/7-1, JA 1890/7-2), Tim Hahn (HA 7070/2-2), Bertram Müller-Myhsok (MU1315/8-2), Astrid Dempfle (DE 1614/3-1, DE 1614/3-2). CP1, biobank: Petra Pfefferle (PF 784/1-1, PF 784/1-2), Harald Renz (RE 737/20-1, 737/20-2). CP2, administration. Tilo Kircher (KI 588/15-1, KI 588/17-1), Udo Dannlowski (DA 1151/6-1), Carsten Konrad (KO 4291/4-1). Data access and responsibility: All PIs take responsibility for the integrity of the respective study data and their components. All authors and coauthors had full access to all study data. The FOR2107 cohort project (WP1) was approved by the Ethics Committees of the Medical Faculties, University of Marburg (AZ: 07/14) and University of Münster (AZ: 2014-422-b-S).

Illness and Treatment Outcome in Depression: A Meta-Analysis. 2011.

Supplementary Information for

# A Network Control Theory Approach to Longitudinal Symptom Dynamics in Major Depressive Disorder

<u>Supplementary Introduction</u>

The general idea of symptom change depending on network typology has also been put forward in NTP: van Borkulo et al. suggested that a stronger connection between symptoms at baseline would hamper remission after two years.[1] While initial evidence seemed to support this claim, the study considered only a single time-point for prediction and used cross-sectional rather than individual symptom-network topology. Also, subsequent replication attempts failed.[2] By treating symptom dynamics analogous to physical systems, we refine the idea in several ways: First, building on the linear time-invariant control system formulation outlined above links symptom-network topology to physical quantities in dynamical systems, thereby providing a comprehensive theoretical basis for the putative relation of symptom change and network-topology. Specifically, we no longer base predictions on overall connection strength, but draw upon *Network Control Theory* to derive how network topology relates to symptom change in each individual (cf. Equation 1 in *Methods*). While, in contrast to [1], *Control Theory* allows for quantitative predictions which differ for different symptom states to be drawn, longitudinal data from each individual is required.

<u>Supplementary Methods</u>

*ReMAP app – Implementation and Recruitment*

The Remote Monitoring Application in Psychiatry (ReMAP) app was developed by researchers of the Department of Psychiatry at the University of Münster. It implicates a high-resolution monitoring of activity via smartphone sensors (geolocation, steps, walking distance, and acceleration) in combination with active self-reports regarding sleep and mood, as well as voice samples. The Beck Depression Inventory (BDI)[3] was made available in the app every two weeks with a random variance of two days. Providing self-reports was optional for participants and not conditional for financial compensation.

The ReMAP app was used as an add-on assessment tool for participants of multiple ongoing longitudinal studies. Studies hosting ReMAP are mostly observational neuroimaging studies that implicate multiple structural and functional MRI assessments as well as genotyping, and the assessment of a variety of clinical variables. Included samples stem from the Marburg/MünsterAffectiveDisorderCohortStudy [4,5], the MünsterNeuroimageCohort [6,7], the TIP cohort, and the SEED 11/19 cohort[8]. All cohorts comprise healthy control (HC) participants, as well as different patient groups, including major depressive disorder participants.

Details on the development and recruitment of ReMAP, as well as on its feasibility, acceptance, and the validity of the self-report measures are published elsewhere.[9,10]

Supplementary Table S1

*Demographic and clinical sample characteristics*

|  | N | M | SD | Range |
|---|---|---|---|---|
| Age | 109 | 40.60 | 13.17 | 18-64 |
| Sex (male/female) | 24/85 |  |  |  |
| BDI baseline | 82 | 18.82 | 12.30 | 0-42 |
| CTQ baseline | 92 | 47.30 | 16.82 | 25-94 |
| #Episodes | 91 | 4.09 | 4.44 | 0-26 |
| Duration of illness (month) | 88 | 35.53 | 44.28 | 0-300 |

*Note.* N=Number of samples with data available, BDI=Beck Depression Inventory, CTQ=Child Trauma Questionnaire

Supplementary Table S2

*Listing of existing substance groups*

|  | N |
|---|---|

| Antidepressants | |
|---|---|
| NaSSA | 8 |
| TCA | 1 |
| NDRI | 5 |
| SSRI | 17 |
| SNRI | 27 |
| MAOI | 3 |
| Other antidepressants | 10 |
| Other Medication | |
| Antipsychotics | 24 |
| Mood Stabilizer | 3 |

Note. Medication data was available for n=98 samples.

NaSSA=Noradrenergic and Specific Serotonergic Antidepressant, TCA=Tricyclic antidepressant, NDRI=Norepinephrine and dopamine reuptake inhibitors, SSRI=selective serotonin reuptake inhibitor, SNRI=serotonin and norepinephrine reuptake inhibitor, MAOI=MAO-inhibitor

*Genotyping and calculation of Polygenic Scores*

Genotyping was conducted using the PsychArray BeadChip, followed by quality control and imputation, as described previously.[11,12] QC and population substructure analyses were performed in PLINK v1.90 [13]. The data were imputed to the 1000 Genomes phase 3 reference panel using SHAPEIT and IMPUTE2.[14–16] Imputed genetic data were available for n=60 individuals. For the calculation of polygenic risk scores (PRS; [17]), SNP weights were estimated using the PRS-CS method [18] with default parameters. This method employs Bayesian regression to infer PRS weights while modeling the local linkage disequilibrium patterns of all SNPs using the EUR super-population of the 1000 Genomes reference panel. The PRS were calculated, using these weights, in PLINK v1.90 on imputed dosage data. As training data, we used summary statistics of genome-wide association studies (GWAS) by the Psychiatric Genomics Consortium (PGC) containing 20,352 cases and 31,358 controls for BD [19] and 51,452 cases and 111,769 controls for MDD (without 23andMe) [20]. The global shrinkage parameter φ was determined automatically (PRS-CS-auto; BD: φ=1.24×10⁻⁴, MDD: φ=1.11×10⁻⁴).

For the calculation of ancestry components, the pairwise identity-by-state matrix of all individuals was calculated on strictly quality-controlled pre-imputation genotype data. Multidimensional scaling MDS analysis was performed on this matrix using the eigendecomposition-based algorithm in PLINK v1.90.

Supplementary Results

Patients (N=109) provided between 8 and 100 (median=15; mean=18) BDI measurements over the course of 48 to 576 days (median=134; mean=186) amount to 2,059 BDI measurements in total. BDI sum scores ranged from 0 to 54 (median=11; mean=13.5). Based on this data, we computed the cross-sectional symptom-network over all BDI measurements. Confirming previous findings of non-random network topology, we show significant interdependence between symptoms across participants and time-points (p<0.05, FDR corrected; Supplementary Figure S1).

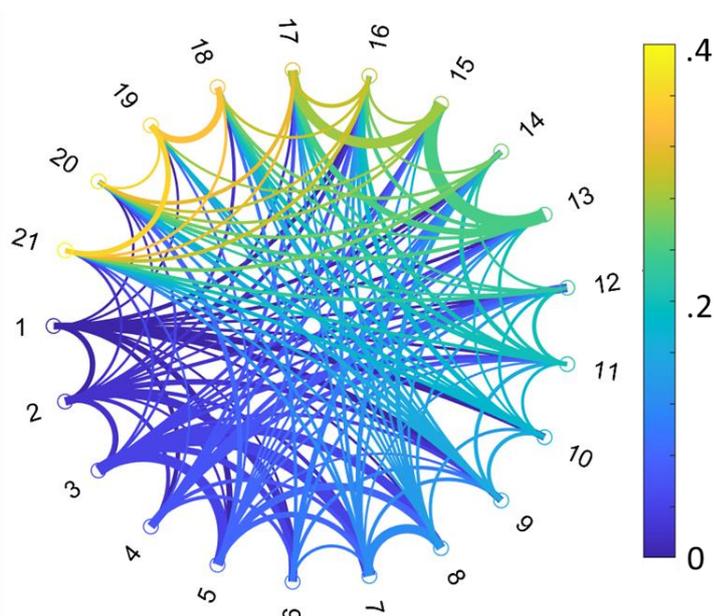

Supplementary Figure S3. Cross-sectional connections between BDI symptoms (p<.05, FDR corrected).

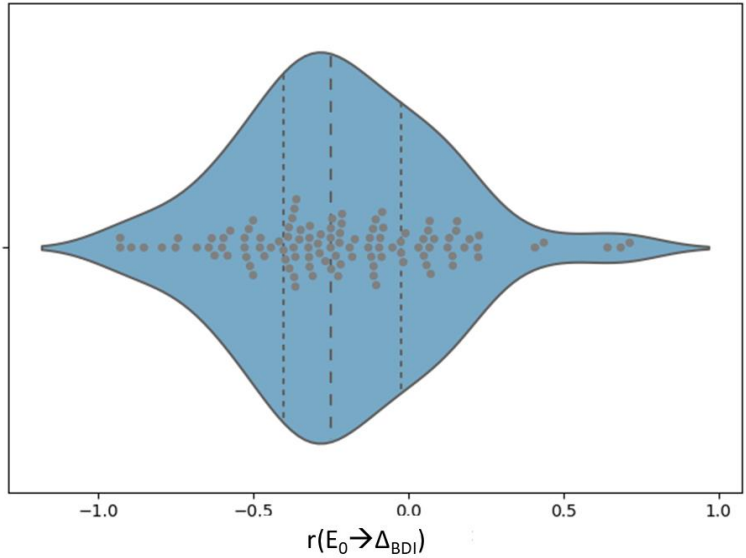

r(E₀→Δ_BDI)

Supplementary Figure S4. Violin scatter plot of the inter-subject variability of the association between $E_0$ at each time-point and BDI improvement at the next time-point ($E_0 \rightarrow \Delta_{BDI}$).

*Driver nodes of complex symptom network*

The question whether and how symptom dynamics can be controlled is of central importance to personalized intervention design. We address this question based on pioneering work of [21] and the generalization offered by [22] known as exact controllably. Specifically, we selected the minimum number of controller nodes also known as driver nodes for each symptom network that satisfies Popov–Belevitch–Hautus (PBH) notion of controllability and thus ensures full controllability. According to the theory of exact controllability, the minimum number of driver nodes ($N_D$) is equal to the maximum geometric multiplicity of eigenvalues of A (also see Equation 1 in the main text). That is:

$$N_D = \ ^{max}_{\ i}\{\mu(\lambda_i)\} \ (3)$$

Where $\mu(\lambda_i) = N - rank(\lambda_i I_N - A)$ and $\lambda_i (i = 1, \ldots, l)$ are the distinct eigenvalues of A. Assuming that the solution to equation (3) is $i = M$, the driver nodes are selected in a way to ensure $[\lambda \quad |M I_N - A, B]$ is full rank (for an efficient optimization algorithm see [22]).

Based on this, we find that number of driver nodes is heavily subject dependent ($N_D \in \{1, 2, ..., 21\}$). However, an average of five symptoms are sufficient (median ($N_D$) = 5)) to control symptom dynamics. The symptoms relevant for the majority of patients pertain to weight-loss, punishment feelings, suicidal ideation, appetite, and interest in social interaction (i.e. items 19, 6, 9, 18, and 12 in the BDI). Importantly, considering individual driver nodes offers a principled way to design personalized interventions which can then be tested empirically.

Supplementary Discussion

Despite the insights and opportunities that may emerge from a *Network Theory of Psychopathology*, numerous obstacles and limitations must be considered.

First and foremost, the approach hinges on a reasonable estimation of symptom-network topology which – in turn – requires longitudinal data capturing the evolution of symptoms over time. To achieve this, we opted for a smartphone-based assessment of symptoms using the well-established BDI within the ReMAP project (for a dedicated validation, see [23]). While this allowed us to conduct one of the largest longitudinal symptom assessments in the literature totaling 2,059 BDIs from 109 patients with an average of 18 BDI measurements spanning a median of 134 days, we have no way to ensure that this database is sufficient. In particular, future studies over longer time-scales should investigate whether how potential non-stationarity of symptom-dynamics affects estimation. In this regard, finding evidence in accordance with the theoretical predictions of *Network Control Theory* increases confidence in our estimates.

Secondly, we used partial correlations based on Kendall's coefficient to estimate symptom-networks. This is largely equivalent to the Gaussian Graphical Models employed in NTP[24], but accounts for the ordinal scale of the BDI items and provides a robust estimate at lower sample sizes.[25] Nonetheless, it does not limit the number of lags and the structure of the adjacency matrix as in optimized in gVAR which has recently become popular in the field.[24] Once larger sample sizes are available, as is the requirement to optimize the larger number of parameters in gVAR, future methodological research ought to focus on improving the estimation of symptom-networks from longitudinal, ordinal-scale data.

Third, and related, the conceptual question of whether symptoms co-occur because of a latent factor or causally influence one another is as fundamental to NTP as it is to our framework. While the former case facilitates network analysis, network centrality measures might be confounded in the latter.[26] While we mitigate this issue by basing our analysis on longitudinal, not cross-sectional data in this work, to estimate the optimal control trajectories, *Network Control Theory* nonetheless benefits from the identification of causal relationships between inputs (interventions) and the states.[27] Thus, while relying on the BDI follows previous successful work by others[28], future studies may consider more comprehensive symptom measurements above and beyond the focus on core symptoms of depression, using e.g. subspace system identification, which might better mirror causal networks (for an overview, see [29]).

Fourth, our approach relies on the simplified noise-free linear discrete-time and time-invariant network model employed in numerous studies.[30–32] Importantly, even if symptoms displayed non-linear dynamics, using this model would be justified as nonlinear behavior may be accurately approximated by linear behavior in physical systems.[33]

While the mechanisms by which risk and resilience affect the degree to which  patient's depends upon his/her symptom-network structure remains to be revealed, this point of view allows us to reconceptualize risk for certain disorders as dependence on symptom-network topologies in which specific patterns of symptoms can more easily form and remain stable (i.e. form an attractor in symptom-space). Likewise, prevention might be understood as favoring symptom-network topologies in which symptoms are not sustainable over longer periods of time. From this point of view, interventions could not only seek to reach low-symptom states, but might also aim to restructure a patient's individual symptom-network topology directly. Whether this is possible or to what extent existing therapies accomplish this already constitutes an exciting question for future research.